\documentclass[acmsmall]{acmart}
\AtBeginDocument{%
  \providecommand\BibTeX{{%
    \normalfont B\kern-0.5em{\scshape i\kern-0.25em b}\kern-0.8em\TeX}}}

\setcopyright{acmcopyright}
\copyrightyear{2025}
\acmYear{2025}
\acmDOI{Pre-Print}

\acmConference[Pre-Print]{}{2024}{}
\acmBooktitle{Pre-Print} 

\usepackage{graphics}
\usepackage[utf8]{inputenc}
\usepackage[T1]{fontenc}
\usepackage{url}
\usepackage{xcolor}

\acmSubmissionID{1582}

\begin{document}



\title[Virtual Agent-Based Communication Skills Training]{Virtual Agent-Based Communication Skills Training to Facilitate Health Persuasion Among Peers}

\author{Farnaz Nouraei}
\email{nouraei.f@northeastern.edu}
\orcid{0002-0406-4410}
\affiliation{%
  \institution{Northeastern University}
  \city{Boston}
  \state{MA}
  \country{USA}
}

\author{Keith Rebello}
\affiliation{%
  \institution{University of Waterloo}
  \city{Waterloo}
  \country{Canada}}

\author{Mina Fallah}
\affiliation{%
  \institution{Northeastern University}
  \city{Boston}
  \country{USA}
}

\author{Prasanth Murali}
\affiliation{%
 \institution{Northeastern University}
 \city{Boston}
 \country{USA}}

\author{Haley Matuszak}
\affiliation{%
  \institution{Northeastern University}
  \city{Boston}
  \country{USA}}

\author{Valerie Jap}
\affiliation{%
  \institution{Northeastern University}
  \city{Boston}
  \country{USA}
  }

\author{Andrea Parker}
\affiliation{%
  \institution{Georgia Institute of Technology}
  \city{Atlanta}
  \country{USA}}

\author{Michael Paasche-Orlow}
\affiliation{%
  \institution{Tufts Medical Center}
  \city{Boston}
  \country{USA}}

\author{Timothy Bickmore}
\affiliation{%
  \institution{Northeastern University}
  \city{Boston}
  \country{USA}}

\renewcommand{\shortauthors}{Nouraei et al.}

\begin{abstract}
Many laypeople are motivated to improve the health behavior of their family or friends but do not know where to start, especially if the health behavior is potentially stigmatizing or controversial. We present an approach that uses virtual agents to coach community-based volunteers in health counseling techniques, such as motivational interviewing, and allows them to practice these skills in role-playing scenarios. We use this approach in a virtual agent-based system to increase COVID-19 vaccination by empowering users to influence their social network. In a between-subjects comparative design study, we test the effects of agent system interactivity and role-playing functionality on counseling outcomes, with participants evaluated by standardized patients and objective judges. We find that all versions are effective at producing peer counselors who score adequately on a standardized measure of counseling competence, and that participants were significantly more satisfied with interactive virtual agents compared to passive viewing of the training material. We discuss design implications for interpersonal skills training systems based on our findings.
 
\end{abstract}
    
\begin{teaserfigure}
\centering
  \includegraphics[width=0.9\textwidth]{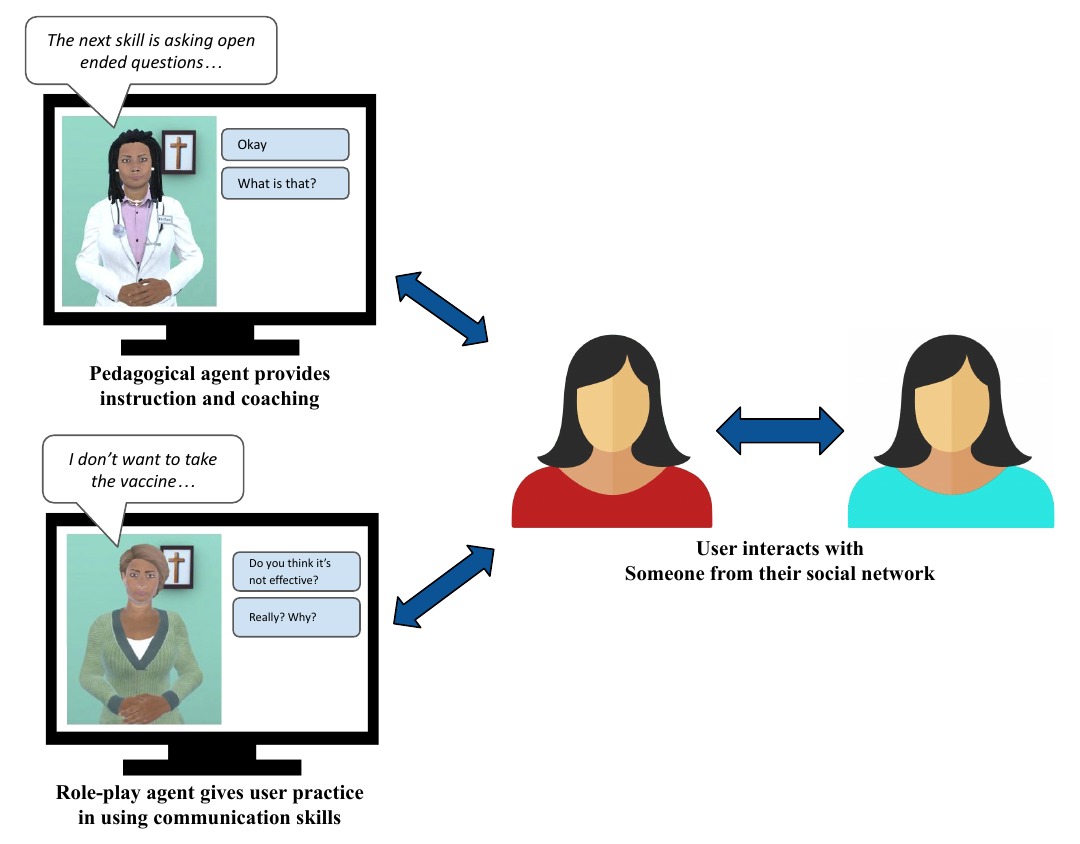}
  \caption{The interaction flow in our web-based lay counselor training  system. A pedagogical agent, Clara (top) and a role-play agent, Mary (bottom) help users with didactic and experiential learning of communication skills, respectively. Users then apply their learning in dialogue with their peers.}
  \Description{Enjoying the baseball game from the third-base
  seats. Ichiro Suzuki preparing to bat.}
  \label{fig:agents}
\end{teaserfigure}


\maketitle
\section{Introduction}
The availability of healthcare workers plays a vital role in ensuring an adequate quality of life worldwide. Unfortunately, many under-served areas continue facing a shortage of health experts, and projections indicate that this trend continues to exacerbate \cite{2019}. Utilizing community-based health volunteers can be one solution, as it has been shown to be effective in increasing the reach of intervention programs while maintaining low cost \cite{walt1990community} in a range of public health initiatives, including managing infectious diseases \cite{lewin2010lay}. 

Although community members are often motivated to encourage their peers for healthy behavior change, counseling involves nontrivial skills of communication. In fact, even when lay counselors receive months of training and refreshers for counseling skills, they may not reach proficiency in delivering care for sensitive topics \cite{dewing2014lay}. Our work was motivated by several health initiatives in which people were provided with communication skills training to influence others in their social network to change health behavior \cite{dewing2014lay, nadkarni2015systematic, giusto2021building}. 

Vaccination is a key example of a health behavior that can benefit from advocacy among community members \cite{2022}. While it remains one of the most effective medical interventions against infectious diseases \cite{decuir2023effectiveness}, vaccine-preventable diseases such as influenza, COVID-19 and human papillomavirus (HPV) kill thousands of people across the globe each year due to low vaccine uptake \cite{tan2015adult}. Increasing awareness of vaccine benefits has been identified as a successful strategy for addressing vaccine hesitancy \cite{jarrett2015strategies}, and face-to-face counseling is most effective for this purpose \cite{thomas2018interventions}. Leveraging peers in interventions to promote vaccination has had some success in increasing its acceptance \cite{gobbo2023peer}. However, conversations about vaccination can be difficult, given the rise of vaccine misinformation in social media  \cite{holt2016importance}. For these reasons, training for specific counseling skills should be provided to community members to effectively engage in these conversations. The goal is to empower lay counselors to offer reliable health information while eliciting and addressing reasons for hesitancy, through face-to-face conversations with those around them \cite{wilhite2022don, miller2012motivational}. 

Traditional approaches to health counselor training often involve the use of standardized patients (SPs), individuals who play the role of a patient to assess counselor trainees \cite{van1990assessment}. The use of SPs in medical school programs has consistently been shown to increase novice providers' confidence and competence in communication and clinical decision-making, through providing a safe environment to practice these skills and opportunities for feedback and self-reflection \cite{flanagan2023standardized, giuffrida2023teaching, hillier2020standardization}. Despite their educational value, however, using SPs can be costly and time-consuming, due to need for cumbersome recruiting, on-boarding, and standardization of their interactions \cite{flanagan2023standardized, whitaker2015motivational, swanson2013assessment}. Automating the counselor training process using virtual agents--animated characters that simulate face-to-face conversations through verbal and non-verbal communication--can reduce this burden \cite{stamer2023artificial}, while ensuring fidelity of the simulated environments to real standardized patient interactions, which has been shown to facilitate learning \cite{jin2014educational}. Owing to their unique communicative abilities, virtual agents have shown promising effects in counseling and healthcare applications \cite{bickmore2007health}, including the training of lay counselors to promote vaccination  \cite{murali2022training}. However, there is not enough insight into why these systems are effective \cite{slovak2015becoming}, especially in the context of vaccination promotion, a gap that we address in our study. Similarly, while previous studies have explored using conversational agents for communication training, many lack empirical evaluations and a focus on persuasive communication in stigma-inducing contexts, such as COVID-19 vaccination \cite{schmid2018future, gavarkovs2019behavioral, tanana2019development, demasi2020multi}.

We describe the design of an intelligent virtual agent system that provides training to users so they can promote vaccination among their peers. We also discuss several design considerations of such systems, including which training curriculum and framework can be effective for lay counselors, and how the communication skills should be presented to a user to maximize satisfaction, understanding, and translation to actual use with individuals in the user's social network. Satisfaction is of importance in this context, since the population targeted by our system, namely lay counselors, are volunteers who choose to learn skills from the system---as opposed to those in a training program who are expected to finish a learning module. It is therefore crucial for our system to engage these volunteers for long-term usage. As a step toward addressing these considerations, we conduct an empirical evaluation of our system to test important features. We focus on promoting COVID-19 vaccination as a case study, motivated by the fact that only half of fully-vaccinated Americans received boosters by the end of 2022 \cite{lin2023vaccinated}, and that COVID-19 vaccination has become a conflictual topic in our society, making it potentially difficult to talk about \cite{link2022}.  

The communication skills taught in our system are techniques from Motivational Interviewing (MI), a brief counseling method for helping people work through indecision and build motivation for healthy change \cite{miller2012motivational}. Our design of the agent system is motivated by research on experiential and active learning, leading to the inclusion of both didactic dialog, in which a pedagogical agent teaches the principles of MI, and role-playing with a different agent, in which the user is given the chance to practice their skills in increasingly difficult scenarios.

\subsection{Motivational Interviewing}
MI is a client-centered counseling technique that revolves around collaboration, empathy, and nonjudgmental communication \cite{miller2012motivational}. The specific communication skills taught by our system include building rapport with small talk, seeking permission to talk about COVID-19, inquiring about vaccination status, asking open-ended questions, active listening, and concluding the conversation by sharing one's own experiences with the COVID-19 vaccine. 
Asking open-ended questions is a cornerstone of counseling, as these questions enable the client to freely express their thoughts and concerns, deepening the counselor's understanding of an individual's perspective and prompting "change talk" – i.e. utterances that indicate a person's willingness to change their behavior.
Active listening entails demonstrating attention to the speaker through verbal and non-verbal behavior, providing reflections (summaries of what the client just said), and conveying empathy, in order to establish rapport and trust.
Our system also encourages respecting the client's autonomy and avoiding unproductive arguments throughout conversations with peers.

\subsection{Experiential and Active Learning}
Agent-based role-playing is a kind of experiential learning, which entails learning through experience and reflection \cite{kolb2014experiential}. According to Kolb's model of experiential learning, a concrete experience can help the learner reflect on observations from their environment and transform them into abstract concepts \cite{kolb2014experiential}. The learner may then experiment with these new concepts in other settings, and continue this cycle by engaging in a new experience. In our system, role-playing with a virtual agent can be considered an experiential component for learning via simulated scenarios that involve a virtual agent with an unvaccinated persona presenting concerns about receiving the vaccine. This component also supports reflection by offering timely feedback based on MI principles.

Active learning---a separate but related concept---focuses on skill development rather than mere transmission of information, and requires that the learner does activities beyond listening, such as engaging in discussions or problem-solving \cite{mckeachie1990research, bonwell1991active}. Discussion methods are effective because students actively process material rather than passively listening, resulting in greater likelihood of elaboration or deep processing and improving retention. The dialogue-based nature of user interactions in a virtual agent system requires users to engage in an activity beyond listening to the lessons presented. Role-playing is another case where active learning is fostered through encouraging higher-order thinking and decision-making \cite{romm1986three}, and through the need for handling human emotions inherent in the scenarios, which can further capture students' interest and engagement \cite{hoover1980college}.  
Many studies have demonstrated the efficacy of experiential learning, showing that students who are provided with multiple opportunities to apply their knowledge and skills can develop their understanding through reflection on success and failure experiences \cite{morris2020, hsu2022effects}. Similarly, many studies have demonstrated the efficacy of active learning strategies in the classroom, such as discussion, over purely lecture-based instruction \cite{mckeachie1990research}. Students are also more satisfied with active learning environments compared to passive learning. However, little is known about the effects of these factors in the context of training for interpersonal skills with virtual agents.

\subsection{Comparative Design Study}
Given this theoretical basis, we conducted a study to evaluate multiple designs for our virtual agent-based lay counselor training system and its ability to prepare users for effective interactions with other people. We intended to address three research questions through this study: 

\textbf{RQ1}. Are virtual agent systems effective at teaching counseling skills and increasing users' self-efficacy in having conversations with community members about vaccination? 

\textbf{RQ2}. How do the effects of dialog-based interaction with virtual agents compare to those of passive learning experiences such as watching educational videos, in terms of student counseling skills and satisfaction, when training for counseling skills? 

\textbf{RQ3}. How might an experiential component, instantiated as the ability to conduct role-playing simulations with a virtual agent, impact student counseling skills and satisfaction in this context?

To investigate these, we conducted a three-treatment randomized between-subjects experiment in which participants used the virtual agent system and then demonstrated their skills with a standardized patient (SP) who played the role of someone who was hesitant to obtain COVID-19 vaccination. The effectiveness of the system is measured from the perspective of the student, student ratings by the SP, and ratings by judges reviewing their recorded counseling conversation. While we demonstrate the use case of COVID-19 vaccination promotion, this training approach may be leveraged for any interpersonal skills training system for behavior change counseling.

We contribute to research in facilitating persuasive interaction in social networks by developing a web-based training system that facilitates peer counseling, by automatically training lay counselors using virtual agents. Through a comparative study, we also provide design contributions to the CSCW community by identifying important design aspects of interpersonal skill training systems that aim to empower peers to have persuasive conversations about health behaviors.

\section{Related Work}
\subsection{Supporting Social Interactions with Technology}
The CSCW literature includes multiple threads of research that focus on technology solutions that help individuals improve their social interactions. For instance, to break awkwardness and encourage organic conversations, \cite{chen2016one} proposed CommonTie, a match-making system that can be used to find peers with similar interests at conferences, by notifying and encouraging the user to engage in face-to-face dialogue with the matched person. Similarly, \cite{kyto2017augmenting} and \cite{hirskyj2019head} explore augmenting interactions with user-curated, digital information about the parties involved, which facilitates conversations between strangers by providing potential knowledge of commonalities. Wearable systems that suggest ice-breaker topics to dyads have also been proposed as a method for tackling the initial discomfort in face-to-face interactions \cite{ogawa2020smartwatch}. Previous research has also investigated the use of smartphone applications that prevent or delay notification interruptions when the user is engaged in face-to-face interactions \cite{park2017don}. While these studies explore facilitating social interactions \textit{in situ} via notifications or information sharing, our proposed system provides training for practical social skills prior to the conversations that a user initiates with others in their community, preparing them for these situations.
\subsection{Social Skill Training Systems}
Several tools have been designed to prepare individuals for social interactions, by providing a space for learning or practicing social skills. In the context of educational games, \cite{jagannath2020we} combined a moderated, multiplayer, online game platform with evidence-based approaches to conflict resolution, to enable adolescents to learn these skills while playing with peers. In their platform, moderators provide support for interpersonal conflicts that naturally arise during the game, and use them as opportunities to teach skills of conflict resolution for the game and real life. Interactive, scenario-based games have also been designed to teach fundamental social and emotional skills such as emotion regulation to children \cite{craig2016enhancing, rubin2011quest, sanchez2014acceptability}. In civics, Diana developed a single-player game that encourages users to engage in productive civil discourse and persuade members of politically polarized communities to focus on their shared values \cite{diana2022persuasion}. Skill-training game environments often require completing short problem-solving challenges in order to advance in the game or reach a goal. Similar to educational gaming platforms, our system includes an experiential learning component where a user can use communication skills to make progress in a persuasive dialogue. Similar to \cite{diana2022persuasion}, we provide training and practice for these skills to a single user, as opposed to involving multiple parties, thus enabling a completely simulated space that allows for inconsequential risk-taking prior to real-life conversations with peers.

Previous studies have employed conversational agents to train individuals for social or interpersonal skills in diverse contexts, such as training police officers for negotiation strategies \cite{bruijnes2013keeping}, negotiation skill training in military scenarios \cite{core2006teaching}, preparing individuals for job interviews \cite{hoque2013mach}, fostering collaborative skills before team activities \cite{samrose2022mia}, and encouraging conversations about health and wellness \cite{albright2016harnessing, murali2022training}, with several of them utilizing role-playing and experiential learning approaches. For example, some prior work has deployed role-playing with conversational technologies, e.g., chatbots, to teach counseling skills in psychotherapy and crisis counseling contexts \cite{demasi2020multi, tanana2019development}, showing general coherence and effectiveness of these tools in learning basic communication skills. Compared to these studies, our work aims to evaluate virtual agents, which involve more than a textual communication modality only, in the context of training lay counselors for skills that contribute to behavior change and vaccination promotion. Schmid Mast et al. explored the potential of immersive virtual reality technology for interpersonal skills training in organizations through role-playing with virtual humans \cite{schmid2018future}, highlighting advantages of this technology such as scalability, flexibility and psychological safety during training. Similarly, Gavarkovs proposed role-playing through virtual simulations for training primary care providers in delivering behavioral counseling \cite{gavarkovs2019behavioral}. However, these studies did not evaluate their proposed approaches in practice, leaving questions about effectiveness and optimal design decisions associated with role-play simulations in virtual environments. Our work, on the other hand, provides a between-subject mixed methods study with an experimental approach to understanding the effects of various design choices, and use of standardized patients to test the trainees' retention of MI skills, leading to a more practical and higher bar for evaluation.

Some previous work has leveraged MI as a technical framework for motivating conversations. For example, Samrose and Hoque created a chatbot system that provides instruction and encouragement for users to develop better communication skills before group collaborations, and found that the agent is more effective when using MI as the main framework for motivating users in changing their current group behaviors \cite{samrose2022mia}. Our training approach differentiates from such methods since our pedagogical agent teaches MI skills to users, as opposed to applying MI in the agent-user interaction. Albright et al. \cite{albright2016harnessing} also used MI as the backbone for their simulated role-play scenarios, to provide experiential training to users on how to initiate dialogue with peers about mental health. In this case, each role-play scenario serves as the main space for learning several skills. Users are also provided with visual guides on how to respond at each turn, and a virtual agent discusses their performance upon completing the role-play. Unlike their work, our system further provides pedagogical training on a skill-by-skill basis, to focus the user's attention on a more concentrated learning scope, immediately followed by practice role-play simulations to maximize retention.

\subsection{Peer Counselor Training and Virtual Agents}

Technology can play an important role in increasing access to peer support, as well as providing opportunities for community volunteers to expand the skills required for delivering such care \cite{o2017design}, especially in low-resource areas \cite{yadav2019leap}. Owing to their potential to enhance healthcare accessibility and efficiency, conversational agents have rapidly evolved as multi-functional platforms increasingly used in healthcare \cite{milne2020effectiveness, tudor2020conversational}. Among these, virtual agents offer unique benefits by providing various communication channels and nonverbal behaviors, which can help establish trust and rapport with users \cite{bickmore2001relational, bickmore2013tinker}. Virtual agents have also been used as a medium for vaccination promotion interventions, leading to higher knowledge and acceptance of vaccines through pedagogical education \cite{pot2017effectiveness} and demonstrating potential to increase user intent to discuss vaccination more effectively compared to traditional approaches \cite{gogoi2022computer}. In addition to traditional pedagogical approaches used in many virtual agent systems for health education, our work incorporates role-play simulations in this context, allowing users to practice communication skills and gain confidence in counseling others in their social network. 

Previous interpersonal skill training systems for vaccination promotion have utilized virtual agents to teach counseling skills to laypersons by means of role-playing \cite{murali2022training}. While this study demonstrated general feasibility and acceptability of the approach, in our current work, we aim to unpack the components of a role-playing training system--including experiential and active learning--and investigate the effect of each on user satisfaction and learning quality. We also evaluate our system in a rigorous manner through the use of Standardized Patients consistently playing the role of vaccine-hesitant individuals, and the use of MI quality measures from clinical psychology.  

\section{A Virtual Agent System for Health Persuasion Skills Training}
We designed a virtual agent system that teaches users counseling skills to promote COVID-19 vaccination in members of their community and others whom they interact with. Given the extensive evidence of the effectiveness of role-playing with both real and virtual standardized patients \cite{flanagan2023standardized, giuffrida2023teaching, hillier2020standardization, stamer2023artificial}, and the merits of experiential approaches to learning \cite{morris2020}, we included a virtual agent in the system that allows users to practice and solidify their communication skills through scenario-based role-playing.

\begin{figure}[t!]
  \centering
    \includegraphics[scale=0.3]{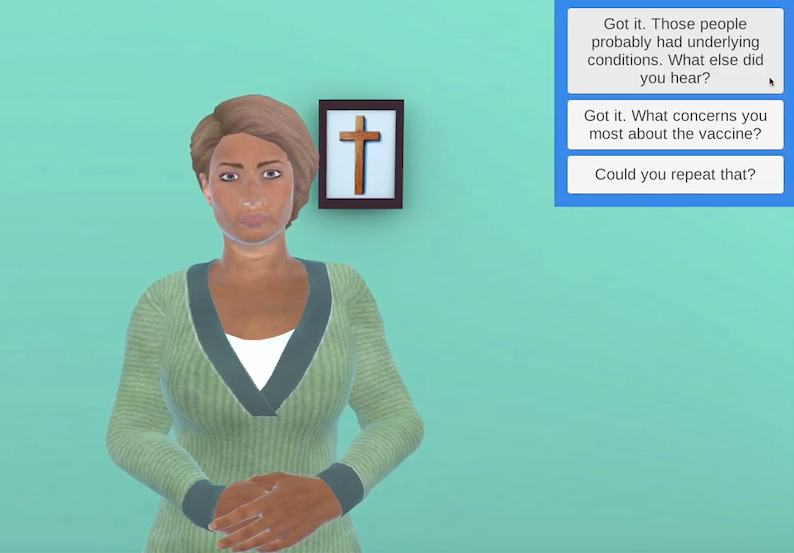}
  \caption{A screenshot of a virtual agent on our web-based platform. This work is part of a larger project, focused on developing a health behavior change intervention for church communities in the US, which led to tailoring visual aspects of the design to these communities.}
  \label{fig:mary}
  \Description{A screenshot of a virtual agent on our web-based platform}
\end{figure}

We developed our system using a web-based framework that provides conversational agents capable of conducting simulated face-to-face dialogue with users. For both agents, conversation is driven in a rule-based manner, using a hierarchical transition network model of discourse based on Grosz and Sidner's theory \cite{rich2001collagen, bickmore2011reusable}. This model represents dialogue as hierarchical layers of finite state machines where each layer represents a discourse segment, designed to attain a particular sub-goal in the conversation (e.g., teaching a user how to ask open-ended questions). Each state within a layer represents an agent utterance, with branching to other states performed following user responses.  We use template-based text generation \cite{reiter1997building} for realizing the natural language representation of user and agent intents. Agent utterances are spoken using a text-to-speech synthesizer (Cereproc\footnote{https://www.cereproc.com}), while user contributions to the dialogue are made via multiple-choice menus, so that all agent responses can be validated for correctness and safety (\autoref{fig:mary}). The dialog content and transition rules based on each user option were created using a custom scripting language, and thus are fully deterministic and involve no statistical approaches. The dialog manager is also connected to a persistent database with information about the intervention and each user, similar to \cite{bickmore2011reusable}, making it possible to scale the dialogue for other interventions, and providing interactions that can be tailored to individual users. Conversational nonverbal behavior for the two 3D animated agents–including hand gestures, posture shifts, and facial display–is generated using BEAT \cite{cassell2001beat}.

We created two agents for our system: pedagogical agent “Clara”,  who provides didactic instruction on counseling skills, and role-play agent “Mary”, who plays the role of a vaccine-hesitant individual that the user can practice their counseling skills with (\autoref{fig:agents}). The agent system's visual design was shaped through iterative community-based participatory design sessions as part of a project to promote wellness in church communities in the US.

The system teaches MI skills in the context of an initial discussion regarding COVID-19 vaccination \cite{miller2012motivational}. Users are taught about an optimal structure of this conversation, the importance of first building rapport, how to introduce the topic of COVID-19 vaccination and how to ask whether someone is vaccinated. They are then taught to ask open-ended questions to understand hesitancy, and do active listening and reflection to demonstrate empathy while maintaining a non-judgmental stance. Finally, they are instructed to share their own experiences regarding vaccination.

\subsection{Training Session Design} \label{section:trainingsess}
When a user begins the session, Clara greets them, gives an overview of the importance of COVID-19 vaccination, and encourages them to learn counseling skills so they can talk to others about the vaccine. Clara then describes the structure of an ideal initial conversation about COVID-19 vaccination and enumerates a set of skills that should be used, presented in the following framework. First, the agent teaches the user how to start a conversation about vaccination by ensuring an anxiety-free environment, rapport-building using small talk, asking for permission for talking about vaccination, and asking whether the client is vaccinated. Second, asking open-ended questions is taught, which allows the client to share their opinions freely without feeling restricted or judged \cite{miller2012motivational}. Third, the user is encouraged to actively listen to what the client is sharing, and respond to them empathically or with reflections, for example, by summarizing what the client has said. This skill not only helps the lay counselor establish a trusting relationship with their client, but also enables them to ask for clarification or follow-up questions and prevent misunderstandings \cite{edwards1997making, miller2012motivational}. Lastly, the importance of sharing the counselor's own reasons for vaccination is discussed, to empower and help clients find commonalities with the lay counselor's story, and connect with them on a more personal level \cite{truong2019role}.

During the pedagogical interaction, Clara teaches each of these skills, followed by introducing the role-play agent Mary as a friend of hers who can help the user practice that skill. Mary plays the role of a middle-aged, healthy woman, who is an acquaintance of the user and is vaccine-hesitant. The user then has a chance to interact with Mary and practice the skill that they just learned. 

An example dialogue flow of interacting with Mary to practice a skill is provided in \autoref{fig:rpflow}. A maximum of two user options are at the disposal of the user at each turn of the role-play conversation, including one response option that adheres to MI principles (e.g., demonstrating active listening by providing a reflection) and one that does not (e.g., dismissing or arguing). While the user options shown in \autoref{fig:rpflow} are marked with terms such as REFLECT and ARGUE for illustration purposes, these descriptive terms are not shown to users during the role-play and they are instead expected to infer these from utterance options, as part of the problem-solving process. If the user fails to choose the right response, Mary immediately provides a context-appropriate response---for example, getting defensive or uncomfortable, and expressing negative feelings---and terminates the interaction by making an excuse to leave. Clara then lets the user know that they made a mistake and gives them another chance to redo the role-play. On the other hand, continuously choosing MI-adherent responses will allow the user to experience the entire role-play for that skill, learn more about Mary's reasoning, and see Mary's positive reactions at the end (\autoref{fig:rpflow}). Once the teaching and practice of one skill are completed, Clara begins to teach the next skill, and the same process is repeated.

The skill-by-skill teaching method used in our training process was designed to help users learn individual skills, and enable them to immediately practice those skills in a simulated role-play scenario, hence resulting in better retention of information through experiential learning. The didactic and role-play components for each skill build upon that of previous skills, and by the end of the training, users will have had the experience of a complete conversation with Mary. If the user passes all skills successfully, this conversation persuades Mary to consider vaccination. Similar to most counselor-standardized patient interactions \cite{wagenschutz2011standardized} and to maintain trainees’ engagement when practicing with Mary, role-play conversations were designed to be short---with an average of 12 turns and about 2 to 10 minutes for each skill, leading to a total of around 25 minutes for the entire system interaction.

\section{COMPARATIVE MEDIA DESIGN STUDY}
We conducted a 3-arm randomized between-subjects experiment to evaluate the experiential and active learning components of our system for training lay counselors. Trainee participants used one of three versions of our training system, which constitute varying degrees of experiential and active learning methods: 

\textit{\textbf{Didactic}}: The user only interacts with Clara, who teaches counseling skills in a step-by-step manner, with no opportunity to practice the skills. This version of the system was designed to manifest active learning, but did not involve experiential learning.  

\textit{\textbf{Role-Play}}: The user interacts with both Clara and Mary, where the former teaches each skill and the latter helps them practice it. This condition was developed to understand the role of role-play practice in the user’s satisfaction and quality of learning. This version used both active learning and experiential learning approaches.   

\textit{\textbf{Video}}: The user passively watches a play-through of the agent system, in which both didactic and role-play components were represented. However, the language in this version is non-tailored (e.g. the user's name is not uttered by the agents), and all components are generic. Additionally, the role-play interactions are success-oriented, showing only what happens when the user makes the correct dialogue move in all situations. Therefore, this version has neither active learning nor experiential learning components, and can be considered a control condition in our experiment.

\paragraph{Standardized Patients (SPs)} To evaluate trainees on their communication skills and adherence to MI principles, we used SPs who played the role of a vaccine-hesitant person that trainees were instructed to encourage for COVID-19 vaccination. Whereas using SPs is a common approach in medicine for both teaching and evaluating students for clinical skills \cite{van1990assessment, flanagan2023standardized, giuffrida2023teaching, hillier2020standardization}, in our study, real SPs were only used for evaluation. Common practices with SPs often involve using one SP with prior experience performing in this role for medical training, in order to maintain consistency across trainee interactions \cite{hillier2020standardization, fink2021assessment}. Due to practical reasons, we recruited two SPs who both had prior experience, and further trained them by explaining the purpose of the study, providing a brief description of MI skills, and explaining the evaluations they would have to perform on trainee participants following each interaction. Both SPs were given a persona (a vaccine-hesitant, middle-aged female) and scenario to practice and were evaluated in practice conversations with the researchers.

Given insights from prior work in lay counselor training, virtual agents, and experiential and active learning, we hypothesized that:

\textbf{H1}: A virtual agent system can successfully train lay counselors to use MI skills to promote COVID-19 vaccination. 

\textbf{H2}: Participants who engage in interactive dialog with the virtual agent system will demonstrate better MI skills in real-life conversations and be more satisfied with their learning experience compared to those who observe a passive recording of the training experience.

\textbf{H3}: Participants who are given the opportunity to practice their MI skills with a virtual agent playing the role of a vaccine-hesitant individual will demonstrate better MI skills in real-life conversations and be more satisfied with their learning experience compared to those who only learn skills through didactic dialog with the agent.

\subsection{Trainee Recruitment}
Trainee participants were recruited using online fliers and university social media platforms, and were required to be 18 years of age or older and up to date on their COVID-19 vaccinations. Even though the system's visual design was tailored to a faith-based intervention due to requirements of a larger project, the current study system did not involve faith-based language and was not limited to faith-based communities, and trainees were recruited from the general public. This study was approved by the Institutional Review Board and participants were compensated \$25 per session.

\subsection{Study Procedure} 
Trainee participants were scheduled for a 90-minute Zoom meeting over which the study would be conducted. Each participant was then randomly assigned to one of the three conditions for the study. A SP was then assigned to each trainee based on their availability.

Once the trainee joined the 90-minute meeting, a baseline survey was administered. Trainees were then given a link to a webpage to use the training system based on their assigned condition. After completing the training, a second survey was administered to assess post-training measures. Following this, trainees were informed that they would be having a conversation with another person, and were given background information about the SP based on the persona. The SP then joined the meeting, and the trainee was given a maximum of 15 minutes to speak with them while the interaction was being recorded by the researcher. The time-bounding of conversations with SPs was done to ensure fair evaluation of trainees' performances while avoiding digressive conversations. After the conversation, both the trainee and SP were given separate surveys. One-on-one semi-structured interviews were then conducted with both the trainee and SP, with questions about their impressions of the system when applicable, and their interaction with the other.

\subsection{Measures}
\autoref{tab:measures} shows the list of measures administered as scale response items at different stages of the study. We also collected two ratio measures post-study from our system, including the time taken to complete training (for all participants) and the number of role-play mistakes (for role-play condition participants). In addition to these, we included post-media manipulation checks in the trainee exit interviews by asking two single questions including: “\textit{How interactive did you feel the training session was?}” (which we refer to as \textit{interactivity}, to measure the manifestation of active learning) and “\textit{How much practice in talking to someone about COVID-19 vaccination did you feel the training gave you?}” (which we refer to as \textit{practice}, to measure the manifestation of experiential learning).

We also had independent judges assess trainees’ communication skills by reviewing video recordings of the interactions. The process of coding followed the MI Treatment Integrity (MITI) coding manual \cite{moyers2016motivational}. First, two key qualities of MI, partnership and empathy, were assessed by giving each trainee interaction a global 1-5 Likert score. Partnership refers to the spirit of treating the client as an equal and collaborating with them to establish a plan for behavior change. Empathy measures the degree to which the counselor attempts to deeply understand the client's perspective and experiences. Partnership and empathy scores were then averaged to represent a Global Relational score. The coders also counted the frequency of different MI-adherent behaviors, such as questions and reflections. To evaluate basic MI competence, the ratio of the number of reflections and the number of questions used by the trainee was calculated (R:Q Ratio). Global Relational scores and R:Q ratios were then compared against MITI thresholds for basic competence in counseling \cite{moyers2016motivational}, and trainees with a minimum score of 3.5 for Global Relational and a minimum of 1 for the R:Q ratio were considered proficient in MI.

Furthermore, video recordings were reviewed for the trainee's usage of specific MI-adherent behaviors that were included in our training material, which include small talk, asking for permission to discuss the vaccine, identifying the client's vaccination status, asking open-ended questions, actively listening, and sharing personal experiences. Each of these skills were rated using a 5-point Likert scale, with anchors that ranged from not showing any of these behaviors to excellent performance.
\begin{table} 
  \caption{Measures used in the comparative study. The first five rows were administered to the trainee only. After the trainee interacted with the standardized patient, the SP rated the trainee’s communication and MI skills, using two measures of CEMI and MI Checklist. The SP and trainee both completed the final three measures.}
  \label{tab:measures}
  \resizebox{0.7\columnwidth}{!}{%
  \begin{tabular}{ccccl}
    \toprule
    Measure Title&Administered&\#Items (Type)&Min Anchor&Max Anchor\\
    \midrule
    Confidence in Counseling &Baseline, After Training& 1 (7-point)&Not at all&Very much\\
    Motivation in Counseling &Baseline, After Training& 1 (7-point)&Not at all&Very much\\
    Satisfaction with Training Media \cite{murali2021friendly} &After Training&11 (7-point)&Not at all&Very much\\
    Satisfaction with Role-Play \cite{murali2022training}&After Training&4 (7-point)&Not at all&Very much\\ 
    Working Alliance (bond subscale) \cite{murali2021friendly, horvath1994working}&After Training&12 (7-point)&Not at all&Very much\\
    Client Evaluation of MI (CEMI) \cite{madson2013measuring}&After SP&16 (4-point)&Never&A great deal\\
    MI Checklist \cite{czart2014using}&After SP&10 (4-point)&Strongly disagree&Strongly agree\\
    Active Empathic Listening \cite{bodie2011active}&After SP&11 (7-point)&Not at all&Very much\\
    Relationship Quality \cite{murali2022training}&After SP&4 (7-point)&Not at all&Very much\\
    Responsiveness of Other \cite{utami2019collaborative}&After SP&4 (7-point)&Not at all&Very much\\
    
  \bottomrule
\end{tabular}
}
\end{table}

\section{Results}
A total of 36 trainee participants were recruited with 12 assigned to each study group. Participants include 13 (36.1\%) Males and 23 (63.9\%) Females. Participant ages ranged from 22 to 69 years, with a median = 27 and IQR = 9.5. All trainees were fully vaccinated for COVID-19. The full list of participant characteristics and their assigned study conditions is shown in \autoref{tab:participants}. Two female actors were recruited to serve as standardized patients, both of ages above 40.

\subsection{Quantitative Results}
Data normality was tested using Shapiro-Wilk tests for all composite and ratio measures. We performed one-sample statistical tests to examine H1, and independent samples tests and correlational analyses were conducted to test H2 and H3. Overall and per-condition descriptive statistics are shown in \autoref{tab:descriptivesandonesample} and \autoref{tab:descriptivesandindependentsamples}, respectively. Significant results are also shown in \autoref{fig:results1} and \autoref{fig:results2}.

\subsubsection{Trainee after Using Training System}

First, to check whether our condition manipulations were effective, we tested the single-item manipulation measures received from trainees, using non-parametric tests. Kruskal-Wallis tests showed that the perceived \textbf{Interactivity} of media was significantly lower in the video condition compared to the other conditions (H(2)=7.1, p<0.05). Pairwise tests after Bonferroni corrections indicated that the video condition was rated significantly lower in interactivity compared to role-play (U=6.2, p<0.05). Moreover, perceived level of \textbf{Practice} was trending higher in role-play than the other two conditions (H(2)=5.2, p=0.07).

To check for effects of demographics, we performed two-way ANOVA and Aligned Rank Transforms for all outcome measures, while taking study condition and age/gender as factors. Main effects of age were found, such that trainees younger than the median age of 27 rated themselves higher than older participants in terms of increase in Confidence in Counseling (Z=2.41, p<0.05) and Working Alliance (t(34)=2.13, p<0.05). For other measures, including MI performance metrics, no significant effects of age were found. Main effects of gender were also found on outcome measures of Global Relational Score (t(34)=2.97,p<0.01), as well as SP-rated Relationship Quality (t(34)=2.28, p<0.05), Active Empathic Listening (t(34)=2.0, p=0.053), Responsiveness (Z=2.17, p<0.05), and MI Checklist (Z=2.36, p<0.05), such that women scored higher compared to men. That said, we found no significant interactions between age/gender and the study conditions for any of our outcome measures, confirming that these characteristics did not contribute to variations in our results.

\begin{figure}[t!]
  \centering
    \includegraphics[width=0.7\textwidth]{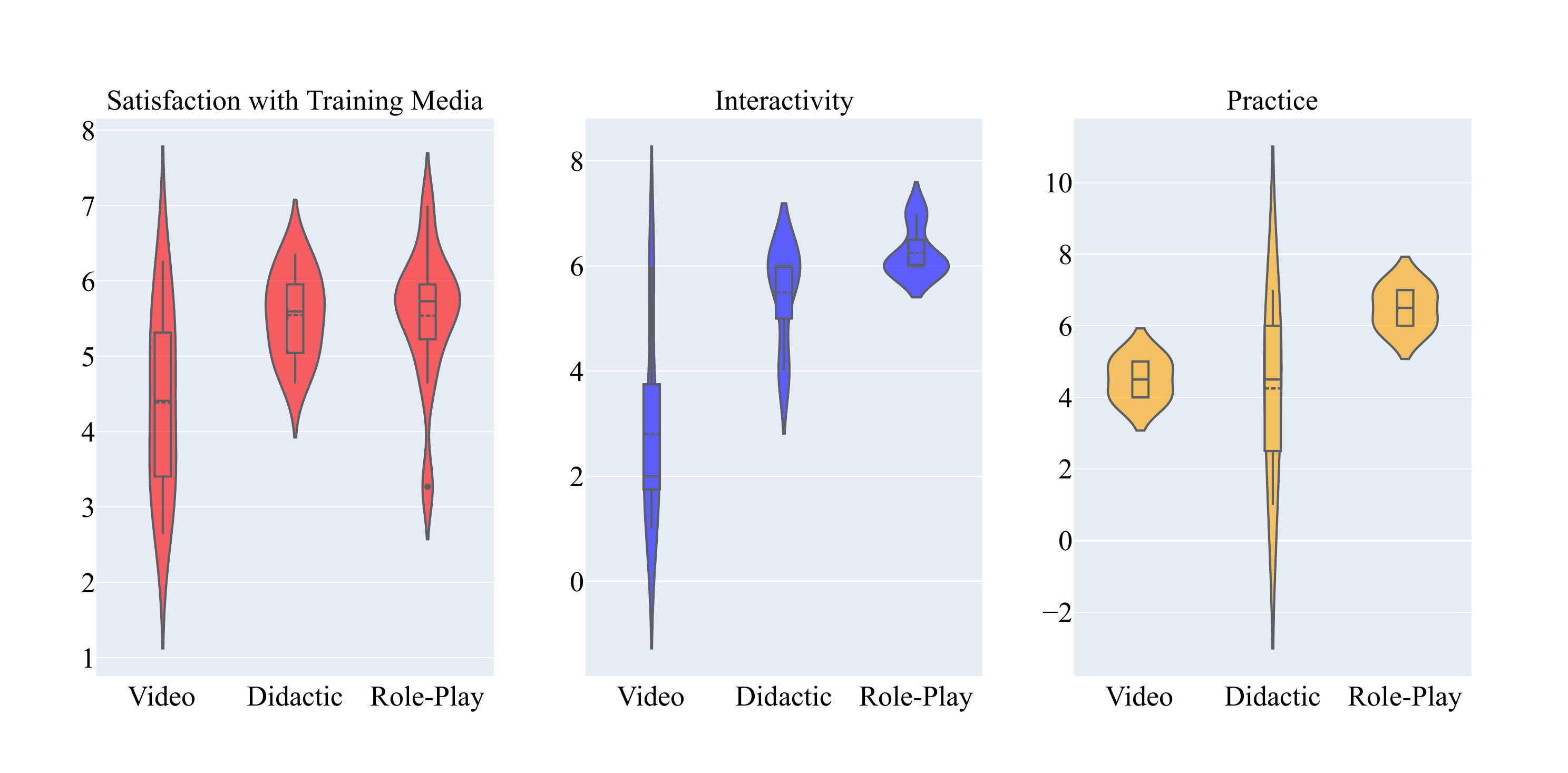}
  \caption{Between-group comparison of Satisfaction with Training Media measure (p<0.05), as well as manipulation checks of Interactivity (p<0.05) and amount of Practice the training provided (p=0.07). Video condition was rated lower than both didactic and role-play in terms of trainee satisfaction, and these ratings were positively correlated with perceptions of interactivity and practice.}
  \label{fig:results1}
  \Description{plots for comparing conditions}
\end{figure}

The three groups were also compared based on their perception of the system. Kruskal-Wallis tests showed significant differences in \textbf{Satisfaction with Training Media}, among the three conditions (H(2)=7.6, p<0.05). Pairwise post-hoc tests showed that the video condition was rated significantly lower than role-play (U=10.7, p<0.05), and was also trending lower than the didactic condition (U=9.8, p=0.06). \autoref{fig:results1} shows the ratings for the two manipulation measures and Satisfaction with Training Media. 

Satisfaction with Training Media was also higher than the neutral of 4 when combining all conditions (M=5.41, Z=4.53, p<.001), and was significantly correlated with perceptions of interactivity (rho=0.8, p <.001), practice (rho=0.65, p<.05), and Satisfaction with Role-play (rho=0.74, p <0.001). These indicate general user satisfaction with all conditions, as well as the existence of relationships between users’ general satisfaction, how involved they felt in the training, and how much they felt they practiced the learned skills using the role-play scenarios.

We also found significant positive correlations between the perceived interactivity and \textbf{Satisfaction with Role-Play} (rho=0.9, p<.001). As the latter scale was measured for both role-play and video conditions, this correlation shows a relationship between interactivity and how satisfied users were when involved in a role-play or watching example conversations with Mary, highlighting the importance of interaction and involvement.

Related-samples Wilcoxon signed-rank tests on the single-item measures of confidence and motivation to counsel on COVID-19 vaccination, across all conditions combined, showed that \textbf{pre-post increase in motivation} was significantly above zero (mean pre=5.69, post=6.19, Z=2.99, p<0.01), and  \textbf{pre-post increase in confidence} was trending above zero (mean pre=6.03, post=6.33, Z=1.93, p=0.05). These results indicate that all three conditions experienced improvements in confidence and motivation to talk to others about COVID-19 vaccination. However, Kruskal-Wallis tests showed no significant difference among the three conditions in terms of improvements in confidence and motivation.

\subsubsection{Standardized Patient Ratings} We performed independent samples t-tests and non-parametric Mann-Whitney tests between the two standardized patients for all SP-rated measures to ensure that the SPs did not rate trainees differently, and found no statistically significant differences. Between-group analyses showed no significant differences among the three conditions for SP ratings. However, one-sample Wilcoxon signed rank tests revealed that \textbf{Active Empathic Listening}, \textbf{Relationship Quality} and \textbf{Responsiveness of Other} were all significantly higher than the neutral score of 4 on these scales for all three conditions combined. Similarly, aggregated ratings (Cronbach's alpha>0.8) of the \textbf{MI Checklist} were significantly higher than the neutral score of 2.5 (Z = 4.39, p<0.001), and \textbf{CEMI} showed significantly higher scores compared to the minimum score of 1 (t(35) = 27.76, p<0.001). These results indicate that trainees in all conditions were competent in basic and MI-adherent communication skills, as perceived by the SP. 

We also compared the time taken for participants in each condition to complete their training with the system, and found significant differences using Kruskal-Wallis tests (H(2)=21.6,p<0.001). Post-hoc tests showed that the didactic condition (M=19.8, SD=1.8) took a significantly shorter time to complete compared to both role-play (M=26.4, SD=4.7; Z=-4.6, p<0.001) and video (fixed time = 21.2; Z=-2.7, p<0.05). This is expected as both role-play and video conditions involved role-play simulations with Mary. The non-significant differences between the time taken to complete role-play and video conditions imply that most participants completed each skill only once. For trainees in the role-play condition, the mean number of mistakes made in the role-play simulations was 2.16 (SD 1.69). Significant, negative correlations between the Active Empathic Listening ratings and the \textbf{number of role-play mistakes} during training (rho=-0.6, p<0.05) implied that trainees may have had similar communication problems with the SP as they did when role-playing with Mary. 

\subsubsection{Video Analysis Results} We performed reliability tests on our scales to calculate the agreement between two coders who independently rated a sample of the video recordings. Inter-rater reliability was adequate, with Intraclass Correlation Coefficients (2-way, mixed, consistency model for averaged scores) ranging from 0.696 to 0.940. One coder proceeded to rate the remaining videos.

One sample t-test showed that the \textbf{Global Relational Score} (the average of Empathy and Partnership scales) was significantly higher than MITI’s basic MI competence threshold of 3.5 across all conditions combined (t(35)=2.1, p<0.05). However, on a granular level, trainees in all conditions combined had a significantly lower \textbf{R:Q Ratio} than MITI's basic competence threshold of 1.0 (M=0.5, Z=-3.5, p<0.001). This shows that most trainees asked more questions than they provided reflective summarizations and clarifications. Out of 36 participants across all groups, 8 had higher Global Relational score and R:Q ratio compared to the respective MITI thresholds, and thus were classified to have basic counseling proficiency. We also found significant correlations between the Global Relational score and pre-post increase in motivation (rho=0.34, p<0.05), indicating that participants who gained higher motivations to counsel via training, were more likely to perform well in counseling the SP.

The six ratings for MI skills included in our training (small talk, asking for permission, etc.) showed high internal consistency (Cronbach’s alpha=0.82) and were aggregated as a composite measure. One-sample Wilcoxon signed-rank tests showed the \textbf{Composite of Skill Ratings} to be significantly above the neutral point of 3 on these scales (M = 3.59, Z = 2.37, p<0.05). Furthermore, significant correlations were found between these skill ratings and pre-post increase in motivation (rho=0.38, p<0.05). No significant between-group differences were found on any of the video ratings.

Finally, to investigate whether the interactivity of the system can lead to better learning outcomes, we also combined both interactive conditions (i.e., didactic and role-play) into one group and compared that with the non-interactive, video condition. We still found no effects of this grouping in any of the learning-related outcomes such as CEMI. Further, no significant correlations were found between learning-related outcomes and Satisfaction with Training Media, confirming no interplay between learning and satisfaction.

\begin{table} 
  \caption{Descriptive statistics and one-sample test results to compare with neutral/threshold points}
  \label{tab:descriptivesandonesample}
  \resizebox{0.6\columnwidth}{!}{%
  \begin{tabular}{cccccl}
    \toprule
    Measure&Mean (SD)&p-value&baseline for comparison\\
    \midrule
    $\Delta$Confidence in Counseling &0.30 (0.89)&0.05&0.0\\
    $\Delta$Motivation in Counseling &0.50 (0.88)&<0.01**&0.0\\
    Satisfaction with Training Media&5.16 (1.05)&<0.001***&4.0\\
    Satisfaction with Role-Play&5.04 (1.25)&<0.001***&4.0\\ 
    Working Alliance (bond subscale)&5.02 (1.01)&<0.001***&4.0\\
    Client Evaluation of MI&2.94 (0.42)&<0.001***&1.0\\
    MI Checklist&3.19 (0.62)&<0.001***&2.5\\
    Active Empathic Listening (Rated by SP)&4.92 (1.52)&<0.01**&4.0\\
    Relationship Quality (Rated by SP)&4.72 (1.69)&<0.05*&4.0\\
    Responsiveness of Other (Rated by SP)&5.24 (1.77)&<0.001***&4.0\\
    Global Relational Score&3.84 (0.97)&<0.001***&3.5\\
    R:Q Ratio&0.53 (0.55)&<0.001***&1.0\\
    Composite of Skill Ratings&3.51 (0.91)&<0.01**&3.0\\
  \bottomrule
\end{tabular}
}
\end{table}

\begin{table} 
  \caption{Per-condition descriptive statistics and between-group hypothesis test results}
  \label{tab:descriptivesandindependentsamples}
  \resizebox{0.7\columnwidth}{!}{%
  \begin{tabular}{ccccccl}
    \toprule
    Measure&Didactic: Mean (SD)&Role-Play: Mean (SD)&Video: Mean (SD)&p-value\\
    \midrule
    Interactivity&5.50 (1.11)&5.03 (1.19)&4.96 (0.75)&<0.05*\\
    Practice&4.25 (2.50)&6.50 (0.58)&4.50 (0.58)&0.07\\
    $\Delta$Confidence in Counseling &0.42 (0.67)&0.00 (0.95)&0.50 (1.00)&n.s.\\
    $\Delta$Motivation in Counseling &0.20 (0.80)&0.42 (1.00)&0.58 (0.90)&n.s.\\
    Satisfaction with Training Media&5.55 (0.56)&5.54 (0.93)&4.39 (1.18)&<0.05*\\
    Satisfaction with Role-Play&NA&5.40 (1.28)&4.69 (1.18)&n.s.\\ 
    Working Alliance (bond subscale)&5.06 (1.11)&5.03 (1.19)&4.96 (0.75)&n.s.\\
    Client Evaluation of MI&2.94(0.45)&2.81(0.40)&2.96(0.48)&n.s.\\
    MI Checklist&3.00 (0.74)&3.24 (0.53)&3.19 (0.69)&n.s.\\
    Active Empathic Listening (Rated by SP)&4.58 (1.54)&5.12 (1.29)&5.09 (1.77)&n.s.\\
    Relationship Quality (Rated by SP)&4.42 (1.85)&4.84 (1.43)&4.92 (1.86)&n.s.\\
    Responsiveness of Other (Rated by SP)&4.96 (1.90)&5.23 (1.84)&5.57 (1.65)&n.s.\\
    Global Relational Score&3.87 (1.05)&3.79 (0.99)&3.87 (0.96)&n.s.\\
    R:Q Ratio&0.59 (0.41)&0.53 (0.62)&0.48 (0.63)&n.s.\\
    Composite of Skill Ratings&3.53 (0.99)&3.54 (0.81)&3.45 (1.00)&n.s.\\
  \bottomrule
\end{tabular}
}
\end{table}
\subsection{Qualitative Results}
Interviews with trainees and SPs were recorded and transcribed, followed by rapid thematic analysis to understand common themes \cite{braun2006using}. Three authors performed open coding, agreed on an initial code-book, and reached an overall unweighted Kappa of 0.84 and 99.55\% agreement on a sample of 8 transcripts (6 trainee and 2 actor interviews) after multiple rounds of analysis. The results reported in this section are based on analysis performed on all transcripts, and the final code-book is presented as Supplementary Material. All qualitative analyses were performed using NVivo 12.

Participants in the interactive conditions described the pedagogical agent, Clara, as \textbf{friendly}, \textbf{well-informed}, and “\textit{very \textbf{humanistic} in terms of her facial expressions}” [P10], which made them develop a “\textit{trust [in] her suggestions}” [P3]. The information presented by the system in all conditioned was deemed effective for \textbf{learning communication skills} and made them \textbf{feel more prepared} to use them in life: "\textit{...this was an active exercise for facing [disagreements]...without people yelling at each other}" [P27]. Some highlighted the usefulness of these interpersonal skills in all contexts, including but not limited to, conversations about COVID-19 vaccination. 

Participants in all conditions expressed that the training helped with their \textbf{confidence} and \textbf{comfort} in having a conversation with the standardized patient since “\textit{I knew what I should expect and how the conversation should go.}”[P9]. Most also mentioned that they were able to use their learnings in their conversation with the SP and that it \textbf{improved the quality} of the conversation and the likelihood of \textbf{persuasion}: “\textit{I might have encouraged her to think a bit about vaccinations}” [P23]. Most who interacted with the agent system mentioned that it may be equal to or more helpful than a human trainer, pointing out that it is always \textbf{clear and concise}: “\textit{With Clara, it’s clear and things are...conveyed as they should, but if it is a person it depends on their ability to convey the information}” [P6]. Some also mentioned that the virtual agent \textbf{never gets tired} of their questions, and has the \textbf{availability} to talk whenever they need to. The agent was also perceived as someone who \textbf{listens} and \textbf{does not judge} them for their mistakes, as a human might: “\textit{If I happened to make a mistake a human might have judged, but Clara does not have that ability to judge}” [P31].

Participants in the role-play condition felt that the role-play simulations gave them an \textbf{opportunity to practice} the skills that they were learning and made them \textbf{aware of how they interact} with other people. Some reported that the \textbf{ability to make mistakes and learn from them} helped to have a better conversation with the standardized patient: “\textit{Whatever mistakes I made in the training session, I avoided them here}” [P30]. Participants believed that the feedback mechanism (Mary expressing she was upset and leaving) was helpful to know when they made a mistake: "\textit{The most helpful part was...getting to different reactions when I choose different tracks of questions}" [P31]. However, some emphasized that they would have appreciated \textbf{more clarity and explanations} of what they had done wrong, through summative feedback by Clara.

The \textbf{interactivity} of the training was also represented as an important factor in how the participants perceived it. In the video condition, users reported that the training was very \textbf{slow-paced} and \textbf{not engaging}, and expressed the need for actual interactions: “\textit{I think I would’ve [liked to] click the buttons on my own}” [P6]. By contrast, users in the role-play and didactic conditions reported feeling that the interactivity allowed them to \textbf{stay engaged} in the training session and “\textit{feel more involved}” [P36] and \textbf{attentive} as they “\textit{had to keep asking and acknowledging what she was saying}” [P36]. Those who engaged in role-play exercises also found that the interactivity gave them a chance to think about how they would respond and \textbf{experiment with different answers}: “\textit{...like pick my own answers, even to just see how that would play out}” [P17]. 

In our interviews with the standardized patients, SPs stated that most of the conversations “\textit{went really well}” and they often described the trainees as “\textit{friendly}” and “\textit{well-spoken}”. For instance, most of the trainees were actively “\textit{\textbf{listening} to [their] ideas and [their] thoughts}”, both by following up with what they said using a question or statement, as well as through their body language. That said, a few trainees were perceived as “\textit{a bit nervous}”, “\textit{not listening}”, or not “\textit{in tune with}” their emotions. Moreover, the SPs reported that some trainees showed \textbf{empathy} during the conversation, by making the SP feel that their points were valid.

\section{Discussion \& Design Implications}
Overall, our study highlights the effectiveness of using virtual agents to train lay counselors for persuasive health communication with others, such as  vaccination promotion. It also points to multiple important design decisions with potential implications for intelligent and automated approaches to communication training for healthcare and beyond.

\subsection{Automated virtual agent systems provide an effective learning environment for interpersonal skill training}
Our results indicate that a virtual agent system can effectively train lay counselors in using MI skills to promote vaccination, while also increasing their confidence and motivation to counsel. The effectiveness of conversational agents in education has previously been shown in various learning contexts \cite{albright2016harnessing, samrose2022mia, demasi2020multi, tanana2019development}, and our work confirmed these findings in the context of behavior change and vaccination promotion through lay counselor training. This finding is also consistent with prior work demonstrating the general efficacy of interactive virtual agents for vaccination promotion training \cite{murali2022training}. We evaluated our approach in the context of COVID-19 vaccination, which is known as a controversial topic in many countries, and confirmed our first hypothesis (H1).

Satisfaction scores with all of our system conditions were high and some trainees expressed a preference for agent-based training over in-person programs, pointing to the consistent clarity of how the pedagogical agent explained MI concepts, and the safe and non-judgmental environment that simulated experiences offer.

Although our study was focused on conversations about COVID-19 vaccination, the generalized usefulness of the system for other types of difficult conversations was highlighted in the interviews, which points to the potential of using agent-based interpersonal skill training in other disciplines where these skills have been identified as essential, such as medicine \cite{barth2011efficacy} and leadership \cite{bono2009survey}.

\subsection{Interactive media engages the learner and leads to higher satisfaction.} Our results point to the importance of interactivity in engaging users and increasing satisfaction with the learning experience. Particularly, participants in the interactive (i.e., didactic and role-play) conditions enjoyed having dialogue with the agent, and user satisfaction was associated with how interactive the training was perceived to be. Users emphasized their involvement as a cause of attentiveness toward training during the interviews. H2 was therefore partially supported. This is an important finding because our system is intended as a standalone educational resource which users can access to learn about counseling skills at their own time (and not as part of a mandatory training program), which necessitates enough motivation for users to engage with the system over time. The finding is also consistent with previous work showing that dialogue with conversational agents is associated with user satisfaction and engagement due to factors such as social presence \cite{schuetzler2020impact, jiang2022ai, lee2020hear}. However, our present work is the first to show these effects when applying dialogue interactions to practical skill transfer in a health promotion setting and in the context of persuasive communication training, such as vaccination promotion.

Despite its impact on satisfaction, the role of interactivity in achieving higher MI scores compared to the video condition was not found. The video condition in our study included a demonstration of base-case-scenario example conversations with the role-play agent, which may have been sufficient for learning the MI skills. Observational learning, as in the case of the video condition, has been shown to be an effective learning strategy \cite{bandura2001social, witte2000meta}. As a relevant finding, in our analysis we also compared interactive and non-interactive versions of the system, which revealed no significant impact of system interactivity on learning outcomes, and there were no observed correlations between user satisfaction and learning. These results may be explained in two ways: First, previous work in interactive systems has shown that factors that lead to higher learning do not necessarily lead to higher satisfaction, especially when learning is to be gained at the cost of higher cognitive load or user burden while interacting with a system \cite{fernando2009automated}. Second, the responses to our satisfaction scale items, which asked about the user's positive attitude toward the media and their intent to continue engaging with the system, can be influenced by various factors such as entertainment, interaction novelty, and usability, beyond the quality of learning.

\subsection{Summative reflections may complement just-in-time feedback mechanisms when learning communication skills through role-play simulations.}
We did not observe any significant differences in learning or using MI skills, or in user satisfaction, between the didactic and role-play conditions (H3 was not supported). Despite that, interviews with trainees indicated that they felt role-playing helped them practice and learn the skills better, and positively influenced their conversation with the SP. Our system provided just-in-time feedback through the role-play agent's negative reactions and termination of the interaction if the trainee made a sub-optimal conversation move. Participants in the role-play condition reported that this feedback mode helped them realize when they made a mistake, but often did not enable them to fully understand the reason for their mistake, leading them to express a desire for summative feedback on their performance at the end of each practice conversation. On the other hand, the didactic condition provided pedagogical training only, which may have served as a low-pressure learning experience leading to adequate learning, possibly explaining why we did not see differences between the didactic and role-play groups. 

Explanation and feedback have been identified as critical aspects of learning in general \cite{burgermaster2017role}, and in particular when designing role-play experiences for skill training \cite{albright2018using, albright2016harnessing}, as demonstrated previously. Research in experiential learning and reflective tutoring has also shown that engaging in reflections after problem-solving fosters meaning-making from experiences \cite{morris2020}, and improves learners' assessments of self and their ability to correct their errors in the future, especially when performed through dialogue \cite{katz2003going, perusso2020contribution}. These findings, along with our present work, indicate that just-in-time feedback may not be sufficient without post-simulation reflections on the trainee's performance, and motivate including summative feedback in future experiential environments with virtual agents.

\section{Limitations And Future Work}
Our study had several important limitations, including the small convenience sample used. Even though our qualitative results point to the possibility of using virtual agent systems to replace costly human training in communication skill training, including a study condition with human trainers would be ideal for further understanding this prospect. Although SPs are widely used in healthcare training and evaluation, they may not reflect the actual dialogue that would occur between trainees and the people they interact with in real life. To ensure a consistent SP experience across participants, we only recruited two female SPs in our study. Including more diverse scenarios and SP characteristics for all participants may validate our results further.

Providing monetary compensation to human participants is common practice across disciplines, including HCI. In our study, we provided minimum wage for participant time only. Despite these measures, we acknowledge that payments may have biased the data collected from participant, yet believe that our comparative results should not have been influenced significantly.

Future work can include more complex simulations in such systems, with more complex agent personas, to provide in-depth training. Making the didactic and role-playing experiences longitudinal, with a persistent state, could increase the impact of the training by following up with users after they tried talking to someone in real life and addressing the specific skill deficits they needed help with. There are also many additional counseling techniques, within MI and beyond, that could be added to the skills curriculum. Finally, real-world evaluation with a larger sample size is crucial for effectively validating technologies that target behavior change \cite{klasnja2011evaluate}. We are currently conducting a longitudinal field study to determine the efficacy of the current agent-based skills training system in improving COVID-19 vaccination rates.

\section{Conclusion}
We demonstrate the feasibility and efficacy of using virtual agents to train lay counselors in health persuasion skills, evaluated in subsequent interactions between trainees and standardized patients playing the role of individuals resistant to change. Lay users were able to achieve scores significantly above the threshold for basic counseling competency on a standardized measure, as scored by judges observing the interactions. In comparing multiple versions of our design, we find that the interactivity provided by simulated face-to-face conversation with virtual agents is preferred to more passive training experiences such as watching a video. We ground our approach in a system that teaches several communication skills from Motivational Interviewing, and use COVID-19 vaccination promotion as our task domain. Our approach represents a new application of virtual agents that enables people to have positive influences on the health of their social networks.


\bibliographystyle{ACM-Reference-Format}
\bibliography{references}

\section{Appendix}
\begin{figure}[t!]
  \centering
    \includegraphics[width=0.65\textwidth, height=0.85\textheight]{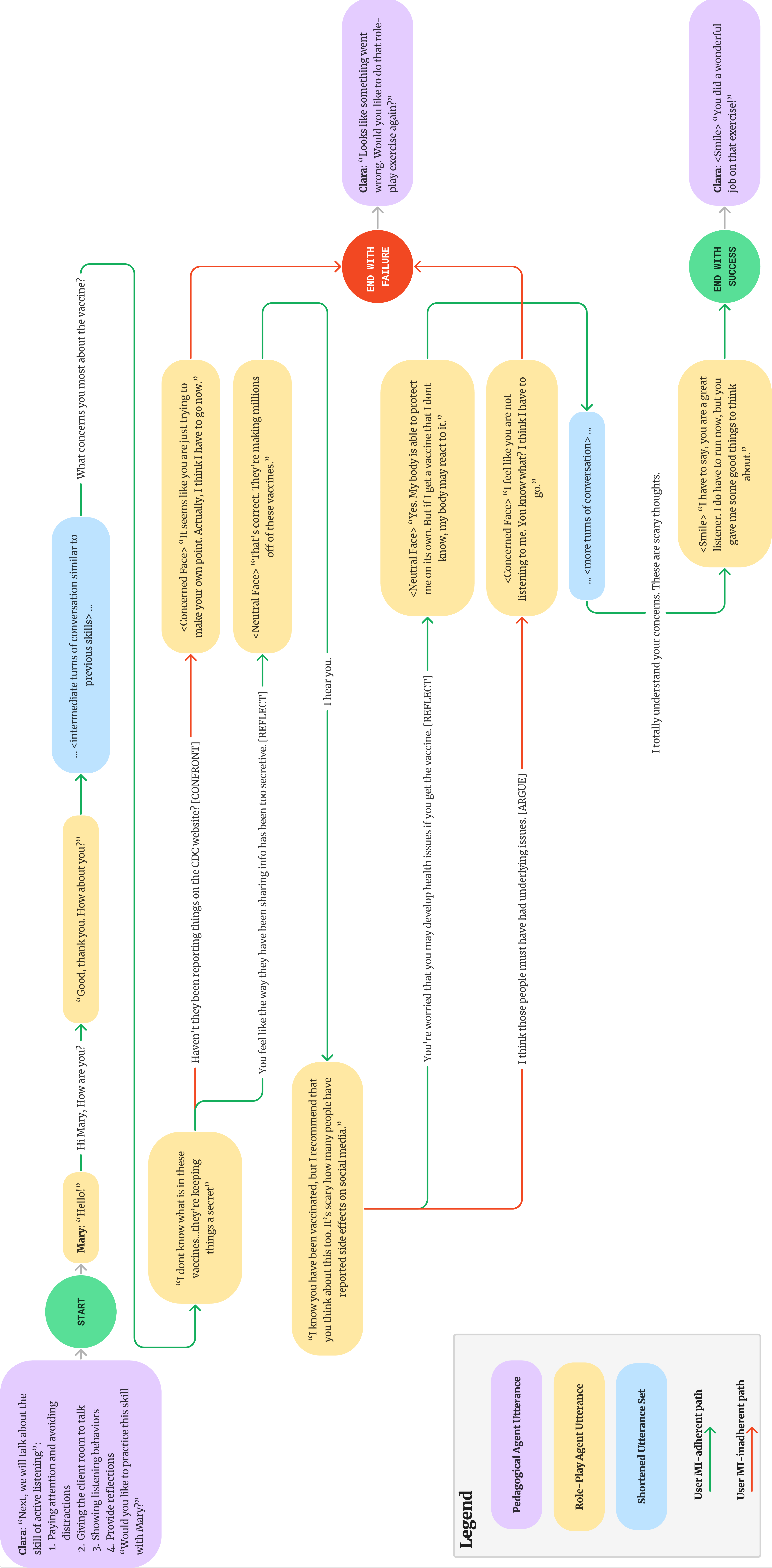}
  \caption{Demonstration of the role-play dialogue flow, using the MI skill of active listening as an example. At each turn, the user is given a maximum of two utterance options that they can choose from, where one option is MI-adherent and one is not. The first blue block shows the omission of some utterances for brevity. These omitted utterances include the segments of the conversation representing the MI-adherent-path utterances from all prior skills, in this case `starting a conversation' and `asking open-ended questions'.}
  \label{fig:rpflow}
  \Description{TODO}
\end{figure}

\begin{figure}[t!]
  \centering
    \includegraphics[width=0.9\textwidth]{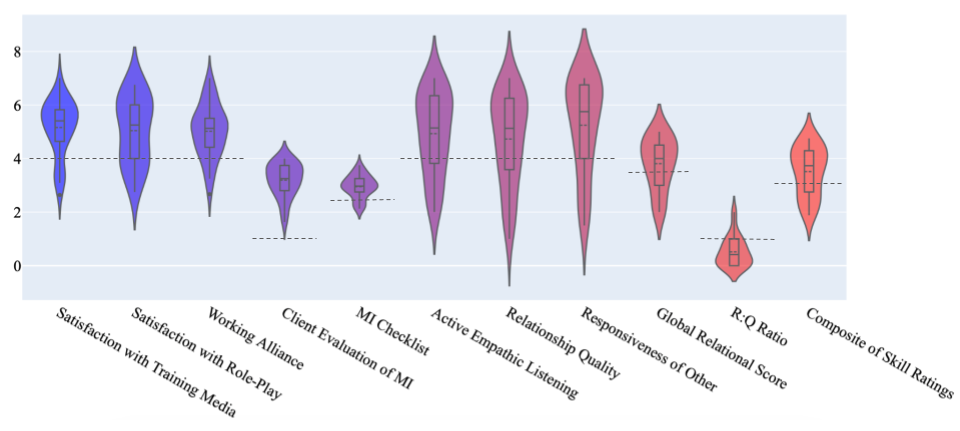}
  \caption{The distributions of scores across all conditions in our study, with dashed lines representing their respective baselines. Measures with a meaningful neutral point in the Likert scale or a MITI-recommended threshold of counseling competency were tested using one-sample hypothesis tests. All Likert measures had means significantly higher than neutral. The mean for Global Relational scores was significantly higher, and the mean for R:Q Ratios was significantly lower than MITI threshold for counseling competency (see \autoref{tab:descriptivesandonesample} for descriptives).}
  \label{fig:results2}
  \Description{plot for scores across all conditions}
\end{figure}

Includes \autoref{fig:rpflow}, \autoref{fig:results2}, and \autoref{tab:participants}.

\begin{table}[]
\caption{List of participant characteristics and their assigned conditions}
  \label{tab:participants}
\resizebox{\columnwidth}{!}{%
\begin{tabular}{lllllll}
\toprule
Participant ID    & Age   & Gender  & Education     & Experience with Computers      & Condition & Condition Type  \\
\midrule
1.00  & 27.00 & Woman   & College Graduate  & Regular User  & Video     & Non-Interactive \\
2.00  & 26.00 & Woman       & College Graduate    & Expert          & Didactic  & Interactive     \\
3.00  & 25.00 & Woman         & College Graduate   & Regular User        & Role-Play & Interactive     \\
4.00  & 26.00 & Woman           & Advanced Degree   & Regular User       & Role-Play & Interactive     \\
5.00  & 25.00 & Woman       & College Graduate & Regular User            & Video     & Non-Interactive \\
6.00  & 25.00 & Man        & College Graduate & Regular User             & Didactic  & Interactive     \\
7.00  & 30.00 & Man            & College Graduate    & Expert            & Didactic  & Interactive     \\
8.00  & 28.00 & Woman             & College Graduate   & Expert          & Role-Play & Interactive     \\
9.00  & 25.00 & Woman            & College Graduate  & Regular User      & Video     & Non-Interactive \\
10.00 & 32.00 & Woman          & Advanced Degree    & Regular User       & Role-Play & Interactive     \\
11.00 & 25.00 & Woman           & College Graduate   & Regular User      & Didactic  & Interactive     \\
12.00 & 22.00 & Man            & College Graduate   & Expert             & Video     & Non-Interactive \\
13.00 & 28.00 & Woman           & College Graduate   & Regular User      & Role-Play & Interactive     \\
14.00 & 24.00 & Man           & Advanced Degree    & Expert              & Didactic  & Interactive     \\
15.00 & 27.00 & Man          & College Graduate   & Regular User         & Role-Play & Interactive     \\
16.00 & 24.00 & Woman             & College Graduate & Regular User      & Didactic  & Interactive     \\
17.00 & 26.00 & Woman          & College Graduate    & Regular User      & Video     & Non-Interactive \\
18.00 & 25.00 & Woman              & College Graduate   & Regular User   & Video     & Non-Interactive \\
19.00 & 33.00 & Man           & Advanced Degree        & Regular User    & Didactic  & Interactive     \\
20.00 & 27.00 & Woman         & Advanced Degree    & Expert              & Video     & Non-Interactive \\
21.00 & 50.00 & Man          & Highschool Graduate   & Regular User      & Role-Play & Interactive     \\
22.00 & 69.00 & Woman          & Some College      & Regular User        & Role-Play & Interactive     \\
23.00 & 26.00 & Man        & Technical School    & Regular User          & Didactic  & Interactive     \\
24.00 & 65.00 & Woman         & Advanced Degree   & Regular User         & Didactic  & Interactive     \\
25.00 & 48.00 & Woman           & Advanced Degree     & Expert           & Video     & Non-Interactive \\
26.00 & 24.00 & Man              & College Graduate    & Expert          & Role-Play & Interactive     \\
27.00 & 63.00 & Man           & College Graduate        & Expert         & Didactic  & Interactive     \\
28.00 & 23.00 & Woman            & Advanced Degree      & Expert         & Video     & Non-Interactive \\
29.00 & 49.00 & Woman             & College Graduate    & Expert         & Video     & Non-Interactive \\
30.00 & 25.00 & Woman              & College Graduate   & Expert         & Role-Play & Interactive     \\
31.00 & 25.00 & Man             & College Graduate    & Regular User     & Role-Play & Interactive     \\
32.00 & 30.00 & Man          & College Graduate    & Expert              & Role-Play & Interactive     \\
33.00 & 65.00 & Woman           & Advanced Degree    & Regular User      & Video     & Non-Interactive \\
34.00 & 35.00 & Woman       & Advanced Degree   & Regular User           & Didactic  & Interactive     \\
35.00 & 47.00 & Man      & College Graduate    & Regular User      & Video     & Non-Interactive \\
36.00 & 29.00 & Woman    & College Graduate    & Regular User         & Didactic  & Interactive \\  
\bottomrule

\end{tabular}
}
\end{table}

\end{document}